\documentclass[epjST]{svjour}
\usepackage{graphicx}
\usepackage{color}

\newcommand {\be}  {
\begin{equation}
}
\newcommand {\ee}  {
\end{equation}
}
\newcommand {\bea} {
\begin{eqnarray}
}
\newcommand {\eea} {
\end{eqnarray}
}

\begin{document}
\title{Elastic and plastic effects on heterogeneous nucleation and nanowire formation}
\author{G. Boussinot\inst{1,2}\fnmsep\thanks{\email{guillaume.boussinot@gmail.com}} \and R. Schulz \inst{2} \and C. H\"uter \inst{2} \and E.~A. Brener \inst{1} \and R. Spatschek \inst{2}}
\institute{Peter Gr\"unberg Institut, 
Forschungszentrum J\"ulich, D-52425 J\"ulich, Germany \and Computational Materials Design Department, 
Max-Planck Institut f\"ur Eisenforschung, D-40237 D\"usseldorf, Germany}
\abstract{
We investigate theoretically the effects of elastic and plastic deformations on heterogeneous nucleation and nanowire formation.
In the first case, the influence of the confinement of the critical nucleus between two parallel misfitting substrates is investigated using scaling arguments.
We present phase diagrams giving the nature of the nucleation regime as a function of the driving force and the degree of confinement.
We complement this analytical study by amplitude equations simulations.
In the second case, the influence of a screw dislocation inside a nanowire on the development of the morphological surface stability of the wire, related to the Rayleigh-Plateau instability, is examined.
Here the screw dislocation provokes a torsion of the wire known as Eshelby twist.
Numerical calculations using the finite element method and the amplitude equations are performed to support analytical investigations.
It is shown that the screw dislocation promotes the Rayleigh-Plateau instability.
} %end of abstract
\maketitle

\section{Introduction}
\label{intro}
The development of elastic and plastic strain is inherent to the problem of assembly of heterostructures, size effects in nanoscopic materials or thermal and mechanical solicitations of macroscopic materials. Since the discovery of strain induced self-assembly of nanowires and -dots \cite{shchukin}, enormous efforts, both theoretical and experimental, have been devoted to the progress in ``strain engineering''. 

In particular, complex geometries of the substrate on which a crystal is deposited are now commonly used to produce highly ordered nanostructures. For example for the semiconductor technology, substrates that are patterned with Si(100) elongated pillars \cite{bergamaschini} are used to produce three-dimensional Ge epitaxial crystals. The resulting structure then depends crucially on the geometry of the substrate and especially on the distance between pillars, from tenth of nanometers to microns. 
In the first part of the article we investigate theoretically the crystallization process in a confined geometry \cite{prl,prb} (for example in the inter-space between such pillars) with a lattice misfit between the crystal and the substrate, focussing on the nucleation regime.
%Here we are thus investigating theoretically, in the first part of the article, the crystallization process in a confined geometry \cite{prl,prb} with a lattice misfit between the crystal and the substrate, focussing on the nucleation regime. 
Whereas the growth and coarsening under the influence of elastic and plastic effects has been investigated already for a long time (see \cite{BrenerMarchenko2000,BrenerHMK2001,BrenerMarchenko2007,BrenerBoussinot2009,FleckerHueter2010,SpatschekEidel2013} and references therein), their role on heterogeneous nucleation is still far less explored.
We use qualitative scaling arguments to produce a phase diagram giving the nature of the nucleation regime as a function of the crystallization driving force and the degree of confinement.
We complement our analytical study with amplitude equations simulations.   

Nanowires, with their almost one-dimensional structure, are intriguing objects from a fundamental point of view, and are promising candidates for future industrial applications on the nanoscale. 
Recently, we have elucidated the interplay between elastic and plastic effects on the equilibrium shape of nanowires \cite{boussinot}.
For any technological application, a long term stability of the nanowire is required and a deterioration through the Rayleigh-Plateau instability \cite{Plateau1873,Rayleigh1878,Karim2006} has been shown to exist as a result of surface energy minimization. On the other hand, already long ago, Eshelby found that a screw dislocation may be stabilized at the center of such a nanowire, leading eventually to a torsion of the whole structure, i.e. the Eshelby twist \cite{Eshelby1953}. Combined with Frank's mechanism for crystal growth from a screw dislocation, striking nanostructures are produced in the form of pine trees \cite{Bierman2008}.
We investigate theoretically the interplay between torsion and surface energy minimization in the second part of the article.

\section{Heterogeneous nucleation with lattice misfit in a confined geometry}
%Crystallization in confined geometries poses fundamental physical questions \cite{prl,prb} and is relevant to a large spectrum of industrial applications. For example for the semi-conductor technology, substrates that are patterned with Si(100) elongated pillars \cite{bergamaschini} are used to produce three-dimensional Ge epitaxial crystals. The resulting structure then depends on the crystallization driving force and on the geometry of the substrate, especially on the distance between pillars.  

The nucleation of a crystalline phase occurs when its free energy density is lower than the one of the vapor (or liquid) phase. In the bulk of the vapor or liquid phase, homogeneous nucleation occurs. If the energy density (energy per unit volume) difference between the phases is $\mu$ and the surface energy of the crystal (energy per unit area) is $\gamma$, the critical nucleus has, according to the classical nucleation theory, a dimension $r^* \sim \gamma/\mu$ and the energy barrier that has to be overcome for the homogeneous nucleation event is $E^* \sim \gamma^3/\mu^2$.

In the presence of a substrate, heterogeneous nucleation occurs on the substrate when the adhesion energy of the nucleated crystal to the substrate is positive. This adhesion energy defines the truncation of the equilibrium shape of the crystal, obtained through the  Wulff construction, taking into account $\gamma_0$, the energy per unit area  of the crystal/substrate interface from which the energy per unit area of the substrate/vapor surface is subtracted. Due to the positive adhesion energy, the energy barrier for heterogeneous nucleation is lower than the energy barrier for homogeneous nucleation. Turnbull showed that their ratio only depends on the wetting angle of the crystal on the substrate \cite{turnbull}. We consider the case where this ratio of the energy barriers is of order unity, i.e. for a wetting angle of order unity. 

Here we first study the scenario where the crystal/substrate interface is coherent. The coherency condition produces an elastic strain in the bulk of the crystal and the bulk of the substrate due to a small misfit $\epsilon$ between their lattice parameters. 
We define a characteristic elastic energy density (energy per unit volume) $e_{el} = Y \epsilon^2$ where $Y$ is an elastic constant. From the interplay between surface energy and bulk elastic energy, a characteristic length scale arises, i.e. the elasto-capillary length $d=\gamma/e_{el}$.
It is reasonable to assume that $d_0 = \gamma_0/e_{el} < d$ (in many cases one even has $d_0 \ll d$). 

At small sizes the misfit strain is relaxed through a continuous change of the shape of the crystal. However, above a critical size, the strain relaxation takes place through the appearance of misfit dislocations \cite{matthews,ertekin,glas}. The characteristic length scale associated to the misfit dislocation mechanism is $b/\epsilon$ where $b$ is the Burgers vector of the dislocations which is of the order of the atomic distance. The system then has a qualitatively different behaviour in the two cases $d \gg b/\epsilon$ and $d \ll b/\epsilon$.

In addition to the introduction of coherency strain effects and their plastic relaxation through the misfit dislocation mechanism, we are investigating the nucleation in a channel of width $H$ between two parallel substrates. This channel may represent the inter-space between pillars on a patterned substrate. $H$ may, in this respect, range from tenth of nanometer to microns and  we therefore consider a continuous macroscopic theory. 
First we study the case where $H$ is infinite.
Then we study the influence of $H$ on the nucleation regime. We are interested finally in producing a diagram that predicts the nucleation mechanism as a function of the two parameters $\mu$ and $H$.

\subsection{The case $d \ll b/\epsilon$: Coherent crystal/substrate interface}
\label{sec:1}

In this section, we investigate the nucleation regime when the misfit dislocation mechanism is inhibited.

{\bf{The case $H \gg d$.}}
 It is known that the bulk elastic energy is efficiently relaxed when the coherent crystal elongates perpendicularly to the substrate \cite{muller}. In the extreme case when the volume of the crystal is much larger than $d^3$, the bulk elastic energy contributes significantly to the energy extremization. The crystal has a cylindrical shape with $r$ being the basal dimension of its interface with the substrate, and $h \gg r$ its height perpendicularly to the substrate. 
In this case, the elastic energy density is of order $e_{el}$ in a region of volume $r^3$ in the neighborhood of the crystal/substrate interface, and vanishes in the rest of the system \cite{boussinot}. The energy of the system may thus be written:
 \be
 E = -\mu r^2 h + \gamma rh + e_{el} r^3 
 \ee
where we neglect the contribution proportional to $r^2$ in the surface energy. The extremization of $E$ gives $r=r_\infty$ and $h=h_\infty$ such that
\bea\label{r_inf}
r_\infty \sim r^* \sim  \frac{e_{el}}{\mu}\; d \;\;\; ; \;\;\;h_\infty \sim r^* \frac{e_{el}}{\mu} \sim \left(\frac{e_{el}}{\mu}\right)^2 d ,
\eea
leading to an energy barrier
$
E_\infty \sim e_{el} d^3 \left(e_{el}/\mu\right)^3 \sim (e_{el}/\mu)  E^* $.
We see that the assumption $r_\infty/h_\infty \ll 1$ requires $\mu/e_{el} \ll 1$. In this regime, the energy barrier for coherent heterogeneous nucleation on the substrate is much larger than the energy barrier for homogeneous nucleation, i.e. $E_\infty \gg E^*$. Heterogeneous nucleation is thus inhibited in this regime. Instead homogeneous nucleation may happen as long as $H \gg r^*$. 
When $d \ll H \ll r^*$, i.e. $1 \ll H/d \ll e_{el}/\mu$, homogeneous nucleation cannot take place and it is likely that the phase transition is inhibited.

In the case $\mu/e_{el} \gg 1$, the bulk elastic energy contributes only slightly to the energetics of the system. Therefore the shape of the critical nucleus is close to the one without elastic effects and the energy barrier for nucleation is only slightly corrected. The reduction of the energy barrier due to the positive adhesion energy is thus sufficient to promote coherent heterogeneous nucleation. A transition between heterogeneous nucleation and homogeneous nucleation therefore occurs when $\mu/e_{el} \sim 1$ in the case $H \gg d$.

%{\it Study of the nucleation on a rigid substrate with amplitude equations. } 
To underline the above predictions for nucleation on a rigid substrate we use the amplitude equations model \cite{spatschek_karma} to simulate the heterogeneous nucleation process with a misfit strain $\epsilon$ for bcc structures.
\begin{figure}
\begin{center}
\includegraphics[width=320pt]{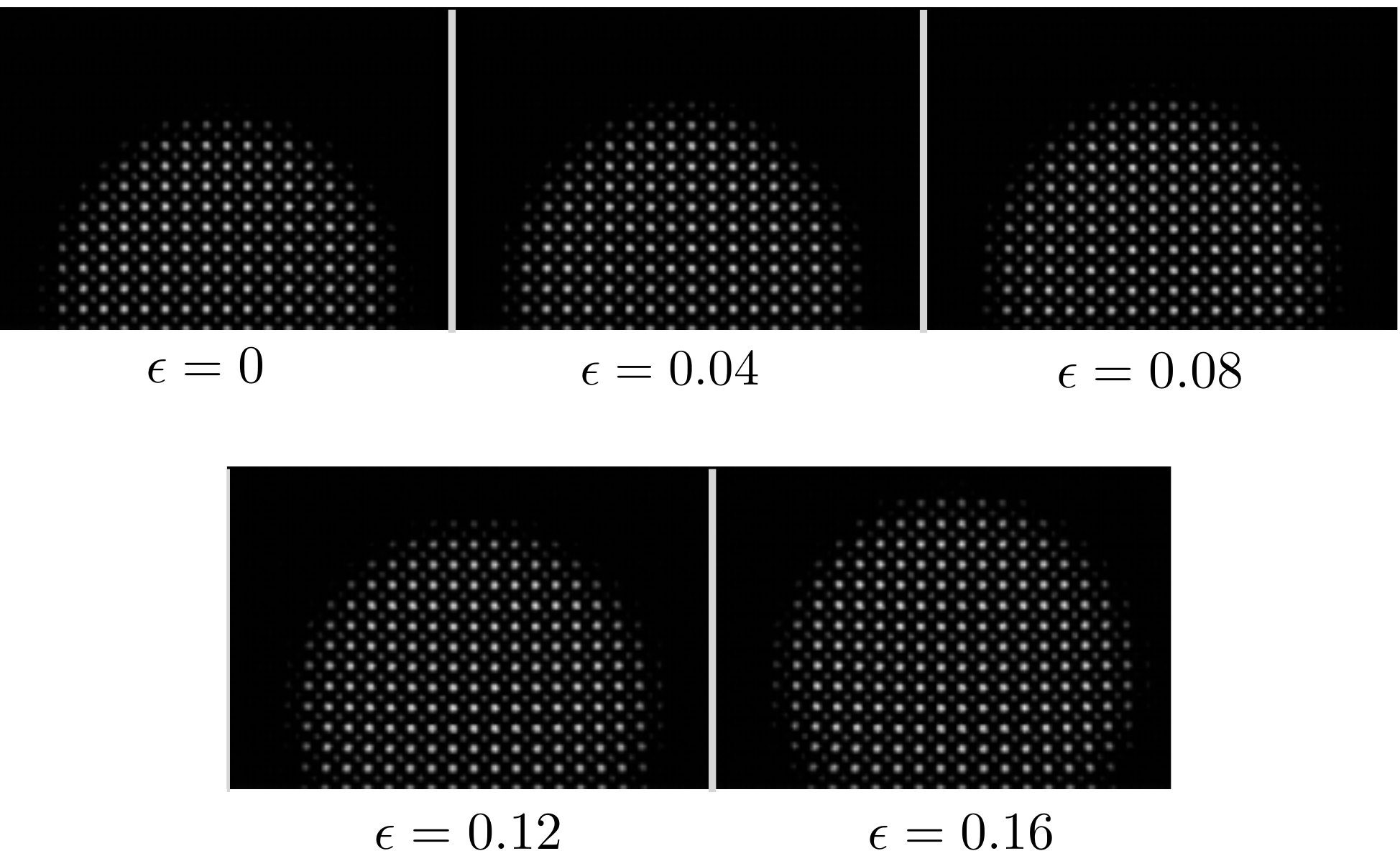}
\caption{Reconstructed atomic density from the amplitude equations for the critical nucleus having different lattice misfit $\epsilon$ with the rigid substrate. The crystal is constrained to have a tensile strain $\epsilon$ at the lower boundary of the simulation box. For each $\epsilon$, only a small part of the simulation box is displayed.}
\label{shapes}       % Give a unique label
\end{center}
\end{figure}
%
%We present now the study of the critical nucleus for different lattice misfits $\epsilon$ using the amplitude equations derived from the two-dimensional Phase-Field-Crystal model developed in \cite{spatschek_karma}. 
In Fig. \ref{shapes} we show the atomic density, reconstructed from the complex amplitudes. We choose a dimensionless undercooling of $(T_M-T)/T_M = 0.0008$, where $T$ is the temperature of the system and $T_M$ is the equilibrium temperature (we refer to Ref. \cite{spatschek_karma} for details of the model). At the crystal/substrate interface (the lower boundary of the simulation box), the crystal is in tension due to a lattice misfit $\epsilon=0, 0.04, 0.08, 0.12, 0.16$ with the rigid substrate.
% is assumed rigid, and its influence on the crystal is provided through boundary conditions for the six complex amplitudes. The moduli of the complex amplitudes obey Von Neumann conditions at the crystal/substrate interface. This allows for the crystal/substrate interface to adjust its size self-consistently. However, the variation of the phase of the complex amplitudes, that is representing the elastic strain, is prescribed according to the lattice misfit $\epsilon$.
Using several initial configurations, the critical nucleus is found, being the one that neither grows nor decays for an extended simulation time. For small $\epsilon$, the critical nucleus assumes a morphology that is close to the one obtained without elastic effects ($\epsilon=0$). When $\epsilon$ increases, the critical nucleus tends to elongate perpendicularly to the substrate, according to the qualitative picture given by Eq. (\ref{r_inf}).
The strain becomes more and more inhomogeneous in the crystal, and for $\epsilon=0.16$ one can clearly see in Fig.~\ref{shapes} the relaxation of the strain with increasing distance from the substrate. 
One should note that for $\epsilon > 0.16$, no critical nucleus is found lying on the substrate, illustrating the transition from heterogeneous nucleation to homogeneous nucleation discussed qualitatively above.

\begin{figure}
\begin{center}
\includegraphics[width=240pt]{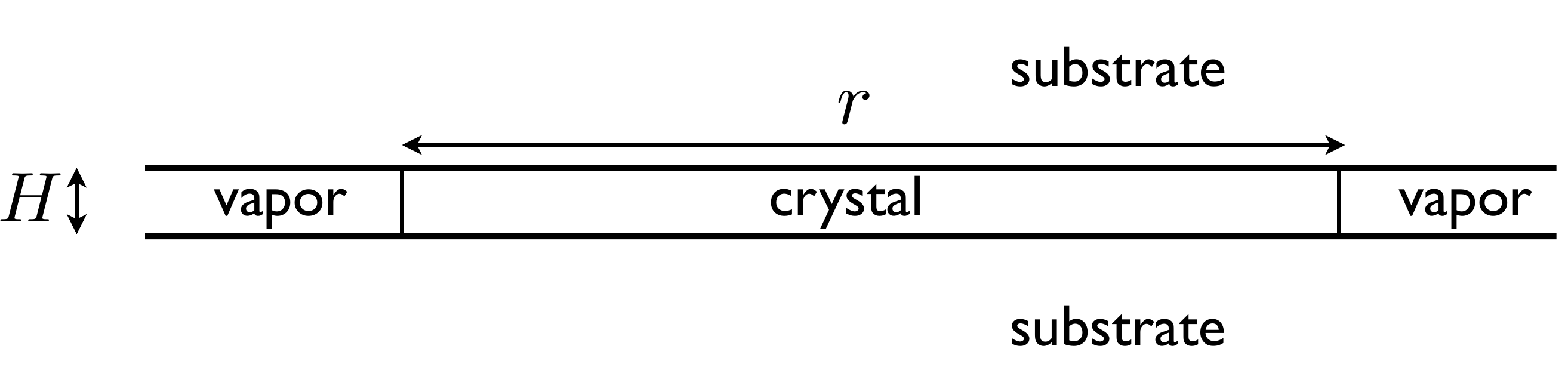}
\caption{\label{schema} Two dimensional cut of the confined nucleus with dimension $r \gg H$.}
\label{fig:1}       % Give a unique label
\end{center}
\end{figure}

{\bf{The case $H \ll d$.}}
 We investigate now a possible confined nucleation where the nucleus is coherently attached to both substrates and has a dimension $r \gg H$. We present in Fig. \ref{schema} a two dimensional cut of this confined nucleus. 
In this confined geometry, the elastic energy density is mainly homogeneous in the crystal and vanishing in the substrate. There exists an intermediate elastic state in a region of volume $rH^2$ in the neighborhood of the crystal/vapor surface (in Fig. \ref{schema} it corresponds to a region of area $H^2$ in the neighborhood of the crystal/vapor surface). We however neglect this contribution.
This assumption is justified in the limit $r/H \to \infty$. 

The energy of the system can thus be written as
\be\label{energy_conf}
E = -\mu r^2 H + \gamma rH + \gamma_0 r^2 + e_{el} r^2 H .
\ee
The extremization gives $r=r_c$ such that $r_c \sim d/x$ 
where $x=\mu/e_{el} - \left( 1 + d_0/H\right)$. The condition $x>0$ provides a threshold for $\mu$, i.e. $\mu > e_{el} + \gamma_0/H$. Here the shift of the transition point is therefore due to elastic effects \cite{brener_JETP} and to the geometrical confinement. 
The dimension of the critical nucleus $r_c$ and the corresponding energy barrier $E_c \sim e_{el} d^2 H/x$ diverge when $x \ll 1$.
Here, the assumption $r_c \gg H$ requires $x \ll d/H$.
Using the assumption $d_0 < d$, we may then connect this confined solution to the heterogeneous nucleation solution when $x \sim \mu/e_{el} \sim d/H \gg 1$, i.e. $r_c \sim r^* \sim H$ and $E_c \sim E^* \sim e_{el} dH^2$.

{\bf{The case $H \sim d$.}}
In this case the heterogeneous nucleation process occurs when $\mu/e_{el} \gg 1$ since $r^* \ll d\sim H$. When $\mu/e_{el} \ll 1$, we have $r^* \gg H$ and the phase transition is inhibited (it may be seen as a continuation of the regime $1 \ll H/d \ll e_{el}/\mu$).
In the situation $H/d \sim \mu/e_{el} \sim 1$ homogeneous, heterogeneous and confined nucleation are competing.
The critical dimensions in these three regimes are of the order $H \sim d$. 

\begin{figure}%[htbp]
\begin{center}
\includegraphics[width=250pt]{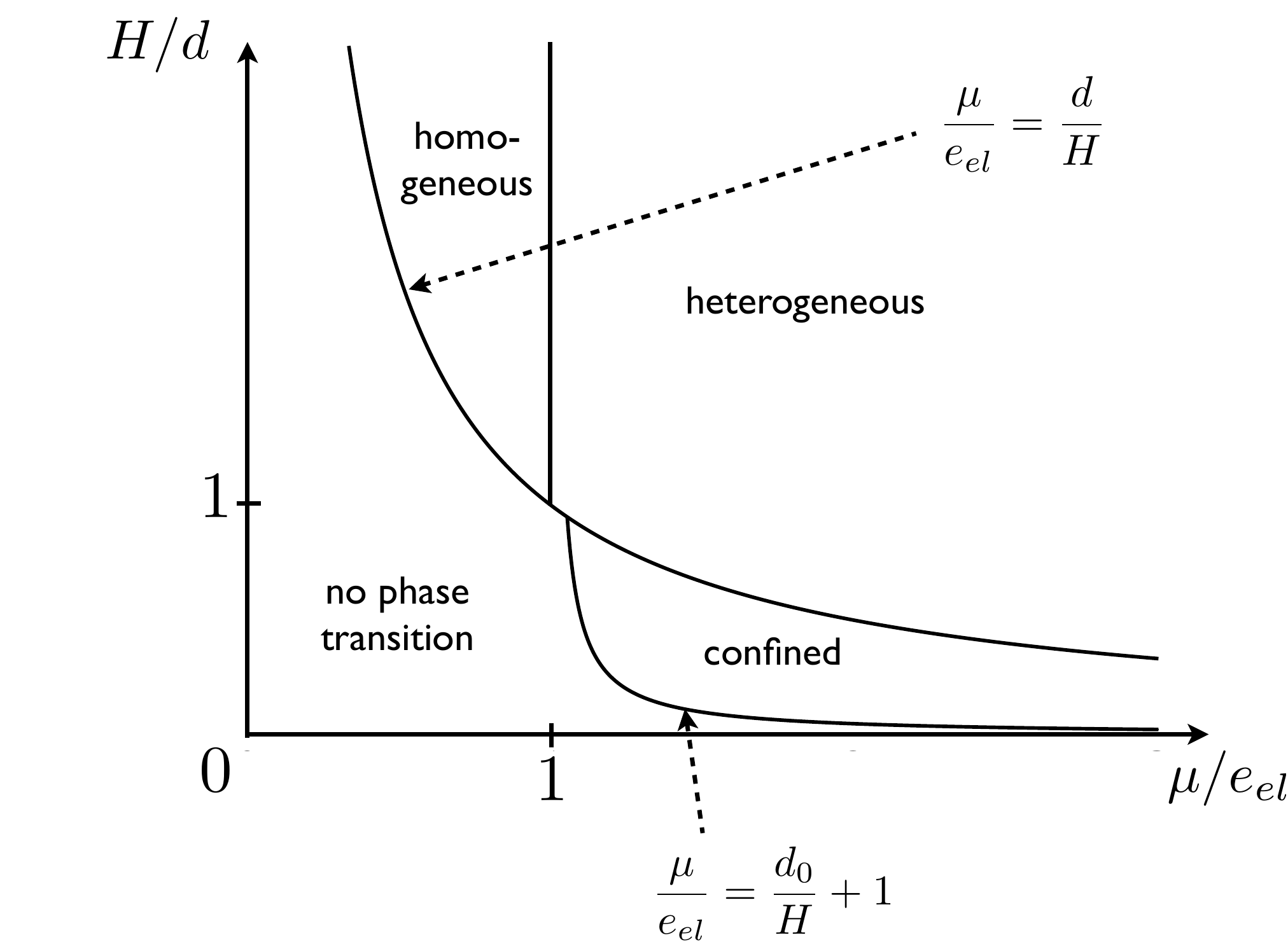}
\caption{\label{diagram} Diagram giving the nature of the nucleation process with respect to the two variables $H/d$ and $\mu/e_{el}$ when no misfit dislocation mechanism is present ($d \ll b/\epsilon$). The details of this diagram in the neighborhood of $H/d \sim \mu/e_{el} \sim 1$ are system specific.
%irrelevant and all regimes are in competition.
}
\end{center}
\end{figure}

%{\bf Summary.}
As a summary we have the different cases listed below, which are represented schematically in Fig. \ref{diagram}:
\bea
1 \ll \frac{e_{el}}{\mu} \ll \frac{H}{d}   &\rightarrow& \mbox{ homogeneous nucleation} \nonumber \\ 
1  \ll \frac{H}{d} \ll \frac{e_{el}}{\mu}  &\rightarrow& \mbox{ no phase transition} \nonumber \\ 
\frac{e_{el}}{\mu}  \ll \frac{H}{d} \ll  1 &\rightarrow& \mbox{ heterogeneous nucleation} \nonumber \\ 
\frac{e_{el}}{\mu} \ll  1  \ll \frac{H}{d} &\rightarrow& \mbox{ heterogeneous nucleation}  \nonumber\\ 
\frac{H}{d} \ll \frac{e_{el}}{\mu} \ll  1 \;\;\mbox{ and } \;\;\frac{\mu}{e_{el}} > 1+\frac{d_0}{H}  \gg 1&\rightarrow& \mbox{ confined nucleation} \nonumber \\ 
\frac{H}{d} \ll \frac{e_{el}}{\mu} \ll  1 \;\;\mbox{ and } \;\; \frac{\mu}{e_{el}} < 1+\frac{d_0}{H}  &\rightarrow& \mbox{ no phase transition} \nonumber 
\eea

\subsection{The case $b/\epsilon \ll d$: Misfit dislocation mechanism}

Another efficient path for a relaxation of the elastic strain is the misfit dislocation mechanism for which dislocations accumulate at the crystal/substrate interface. The length scale associated with this mechanism is $b/\epsilon$ where $b$ is the Burgers vector of the dislocations. When the smallest dimension of the crystal is larger than $b/\epsilon$, a regular array of misfit dislocations builds up and no elastic energy density exists at distances larger than $b/\epsilon$ from the crystal/substrate interface. On the other hand, the energy per unit area of the crystal/substrate interface effectively increases by an amount of order $Yb\epsilon \sim e_{el} b/\epsilon$.
%This length scale describes the decay of the elastic field perpendicularly to the crystal/substrate interface. Two cases should be considered:

%1) When the height of the crystal $h$ is much larger than the dimension $r$ of the crystal/substrate interface, the system lowers its energy by covering the crystal/substrate interface with a regular array of misfit dislocations when $r$ becomes larger than $b/\epsilon$.

%2) When the height of the crystal $h$ is much smaller than the dimension $r$ of the crystal/substrate interface, the system lowers its energy by covering the crystal/substrate interface with a regular array of misfit dislocations when $h$ becomes larger than $b/\epsilon$.

%Thus, the dislocation array minimizes the energy of the system when the smallest dimension of the crystal is larger than $b/\epsilon$.

When $b/\epsilon \gg d$, i.e. $Yb\epsilon \gg \gamma$, the analysis presented in the previous section is relevant because the radius of the critical nucleus $r^*$ is of order $b/\epsilon$ when $\mu/e_{el} \sim d/(b/\epsilon) \ll 1$. Since homogeneous nucleation is promoted in this regime, the dislocation mechanism does not play any role.
In the limit $\epsilon \ll 1$, we have the opposite case $d \gg b/\epsilon$. The diagram giving the nucleation regime as a function of $\mu$ and $H$ is then qualitatively changed. 
However, no changes exist if $H \ll b/\epsilon$ or $r^* \ll b/\epsilon$, i.e $\mu/e_{el} \gg d/(b/\epsilon)$ because the smallest dimension of the crystal is then smaller than $b/\epsilon$.

{\bf Incoherent heterogeneous nucleation.} 
When $H \gg r^* \gg b/\epsilon$, i.e. $d/H \ll \mu/e_{el} \ll d/(b/\epsilon)$, incoherent heterogeneous nucleation occurs. Since, in this case, the energetic cost of the dislocation array is much smaller than the surface energy of the crystal, the dimension of the critical nucleus and the energy barrier change only slightly in comparison to the coherent heterogeneous case. Therefore, when $\mu/e_{el}$ increases, there is a transition from the incoherent heterogeneous nucleation to the coherent heterogeneous nucleation for $\mu/e_{el} \sim d/(b/\epsilon)$, i.e. $r^* \sim b/\epsilon$.

{\bf Incoherent confined nucleation.} 
When $r^* \gg H \gg  b/\epsilon$, incoherent heterogeneous nucleation is not possible. Instead, incoherent confined nucleation occurs.
The incoherent confined nucleus is described energetically by Eq. (\ref{energy_conf}) where $e_{el}$ is set to 0 and $\gamma_0$ is replaced by $\gamma_0 + Yb\epsilon$ where $Yb\epsilon$ is the energetic cost per unit area of the misfit dislocation array. The threshold for $\mu$ then reads  $x_{ic}=\mu/e_{el} - (d_0+b/\epsilon)/H >0$. The critical radius reads $r_{ic} \sim d/x_{ic}$
and the energy barrier becomes $E_{ic} \sim e_{el} d^2 H/x_{ic}$, i.e. $r_{ic} \sim \gamma/[\mu-(\gamma_0+Yb\epsilon)/H]$ and $E_{ic}\sim \gamma^2 H/[\mu-(\gamma_0+Yb\epsilon)/H]$.

By decreasing $H$, a transition from the incoherent confined nucleation to the coherent confined nucleation occurs when $H \sim b/\epsilon$. This transition takes place in the range $1+d_0/(b/\epsilon) < \mu/e_{el} < d/(b/\epsilon)$.

\begin{figure}%[htbp]
\begin{center}
\includegraphics[width=250pt]{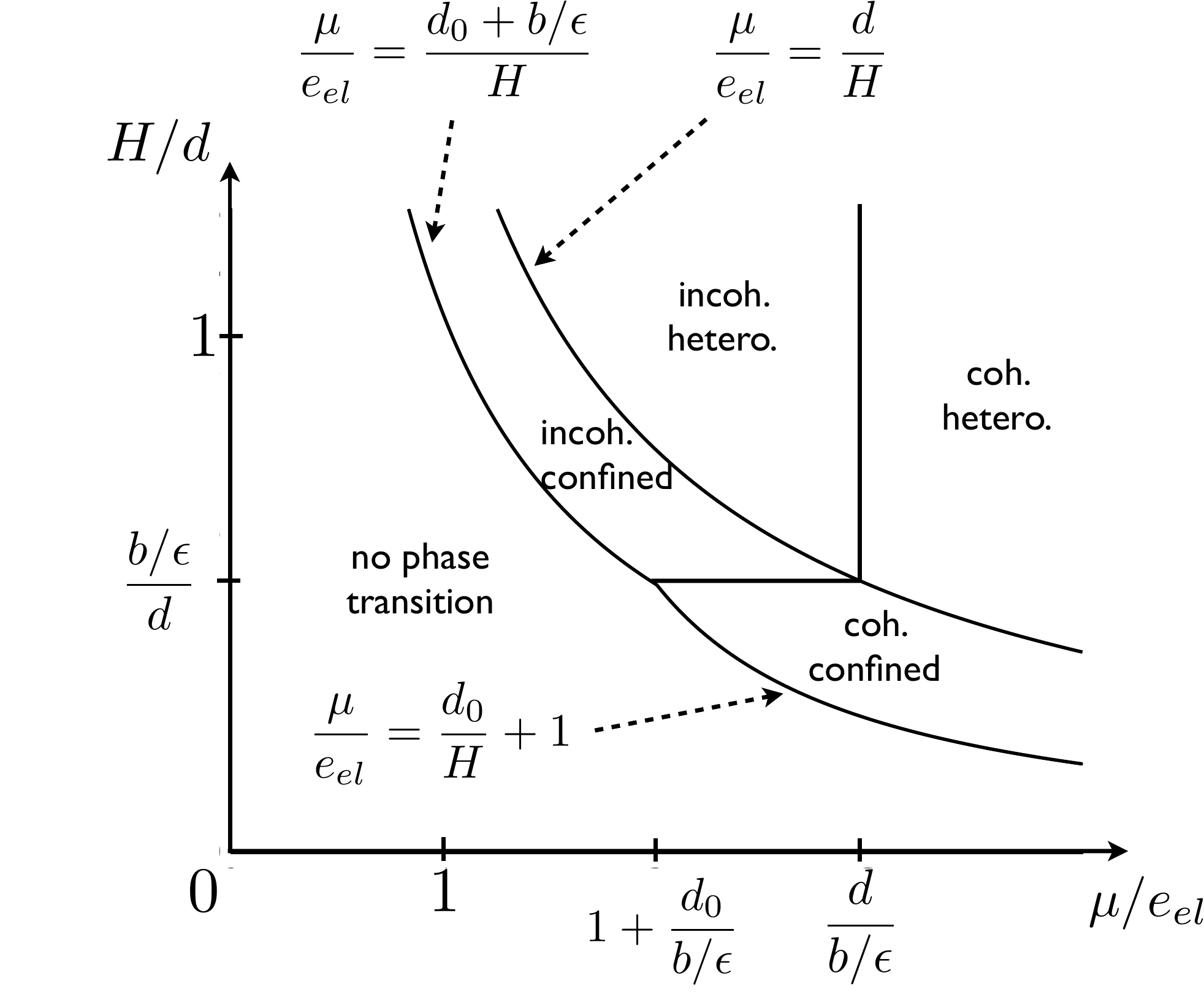}
\caption{\label{diagram_inc_1} Diagram giving the nature of the nucleation process with respect to the two variables $H/d$ and $\mu/e_{el}$ when $b/\epsilon \ll d$.}
\end{center}
\end{figure}

%{\bf Summary.}
We summarize the different nucleation regimes as listed below in the case where $b/\epsilon \ll d$. Two cases should be considered:
\begin{enumerate}
\item
$H \ll b/\epsilon$
\bea
\mu/e_{el} \gg d/H &\to& \mbox{ coherent heterogeneous nucleation} \nonumber \\
1+ d_0/H < \mu/e_{el} \ll d/H &\to& \mbox{ coherent confined nucleation}  \nonumber \\
\mu/e_{el} < 1+ d_0/H  &\to& \mbox{ no phase transition}  \nonumber 
\eea
\item
$H \gg b/\epsilon$
\bea
\mu/e_{el} \gg d/(b/\epsilon) &\to& \mbox{ coherent heterogeneous nucleation} \nonumber \\
d/H \ll \mu/e_{el} \ll d/(b/\epsilon) &\to& \mbox{ incoherent heterogeneous nucleation} \nonumber \\
(d_0+b/\epsilon)/H <  \mu/e_{el} \ll d/H &\to& \mbox{ incoherent confined nucleation} \nonumber \\
\mu/e_{el} < (d_0+b/\epsilon)/H  &\to&  \mbox{ no phase transition}  \nonumber 
\eea
\end{enumerate}
These different regimes are represented schematically in Fig. \ref{diagram_inc_1}.

\subsection{Conclusion}
We have studied theoretically the influence of lattice misfit $\epsilon$ and of the confinement in a channel of width $H$ on the heterogeneous nucleation process. We have used scaling arguments to draw a phase diagram giving the nucleation regime as a function of $\mu/e_{el}$, the ratio of driving force $\mu$ and elastic energy density $e_{el}$, and $H$. Two regimes, with their corresponding phase diagram, have to be considered whether the misfit dislocation mechanism takes place or not.
% The misfit dislocation mechanism is inhibited (takes place) when its characteristic length scale is larger (smaller) than the the elasto-capillary length $d=\gamma/e_{el}$ where $\gamma$ is the magnitude of the surface energy of the crystal. 
%Two different phase diagrams are thus given depending on these two regimes.

In addition to the analytical qualitative investigations we have studied the critical nucleus when the height $H$ is infinite, using amplitude equations. The morphology of the critical nucleus agrees well with the qualitative analytical statements obtained using scaling laws, i.e. the critical nucleus elongates perpendicularly to the substrate when $\epsilon$ increases. Moreover, we found the transition from heterogeneous to homogeneous nucleation around $\epsilon=0.16$. 
Beyond this point we are in the regime where the surface energy $\gamma$ is small enough to prohibit the dislocation entrance.

\section{Eshelby twist and Rayleigh-Plateau instability in nanowires}

Nanowires are structures with a typical diameter of less than 100 nm and a large aspect ratio from length to diameter of more than about 100.
These nanostructures can develop based on heterogeneous nucleation processes at a substrate, usually accelerated by catalytic processes.
Usually, they are produced via vapor-liquid-solid (VLS) growth, where a liquid catalyst is placed on a substrate, on which the nanowire grows from a supersaturated gas phase.
The gas atoms are absorbed by the interface layer between solid and melt, from which the growth proceeds.
The diameter of the nanowires is determined by the droplet size of the liquid catalyst.
Nanowires have many fascinating electronic and mechanical properties due to their almost one-dimensional structure.
Nanowires belong to the best controlled structures on the nanoscale.
As combination of p- and n-doped types they are promising candidates for various semiconductor applications like field effect transistors, diodes, LEDs, complex logical gates, lasers, sensors or solar cells \cite{Lieber2007,Bierman2009}.

For any technological application the long term stability of nanostructures is essential to guarantee functionality and to prevent the generation of ``nano-hazard''.
Here in particular the large surface-to-volume ratio can play an important role.
The reduction of surface energy can lead to morphological instabilities of the wire, which finally leads to the decomposition into droplets for long wave-perturbations, known as Rayleigh-Plateau instability \cite{Plateau1873,Rayleigh1878}.
This has been demonstrated experimentally for gold nanowires, which show the decomposition at rather low temperatures below 500$^\circ$C within a few hours \cite{Karim2006}.
On the other hand, this a priori detrimental development of an instability may be used on purpose e.g.~for thermoelectric purposes.
Here it has been noticed that the thermal conductivity can be reduced by a factor of about 100 for nanowires with a rough instead of a flat surface, which increases the thermoelectric efficiency substantially \cite{Martin2010}.

In this respect, pine tree nanowires may appear as an interesting nanostructure.
They have the special feature that they grow around an axial screw dislocation \cite{Bierman2008}.
By the adjustment of the gas partial pressures during the VLS growth pine tree nanowires can be grown in a very controlled way.
The presence of the screw dislocation induces a torsion of the wire, known as Eshelby twist, which stabilizes the defect in the center of the wire \cite{Eshelby1953}.
Since the dislocation additionally induces elastic stresses, a morphological instability can be substantially influenced by elastic effects \cite{Gugenberger2008,SpatschekFleck2007,SpatschekHartmann2006,BrenerSpatschek2003,SpatschekBrener2001}.
The combination of the Eshelby twist and surface stability is the subject of the following theoretical investigations.

%%%%%%%%%%%%%%%%%%%%%%%%%
\subsection{Eshelby twist}

In the following we assume isotropic linear elasticity, which means that stress and strain are related by Hooke's law, $\sigma_{ij} = 2\mu \epsilon_{ij} + \lambda \delta_{ij} \epsilon_{kk}$.
The strain $\epsilon_{ij}$ is obtained from the displacement field $u_i$ via $\epsilon_{ij}=(\partial_i u_j + \partial_j u_i)/2$.

\begin{figure}
\begin{center}
\includegraphics[trim=1cm 9cm 0cm 0cm, clip=true, width=10cm]{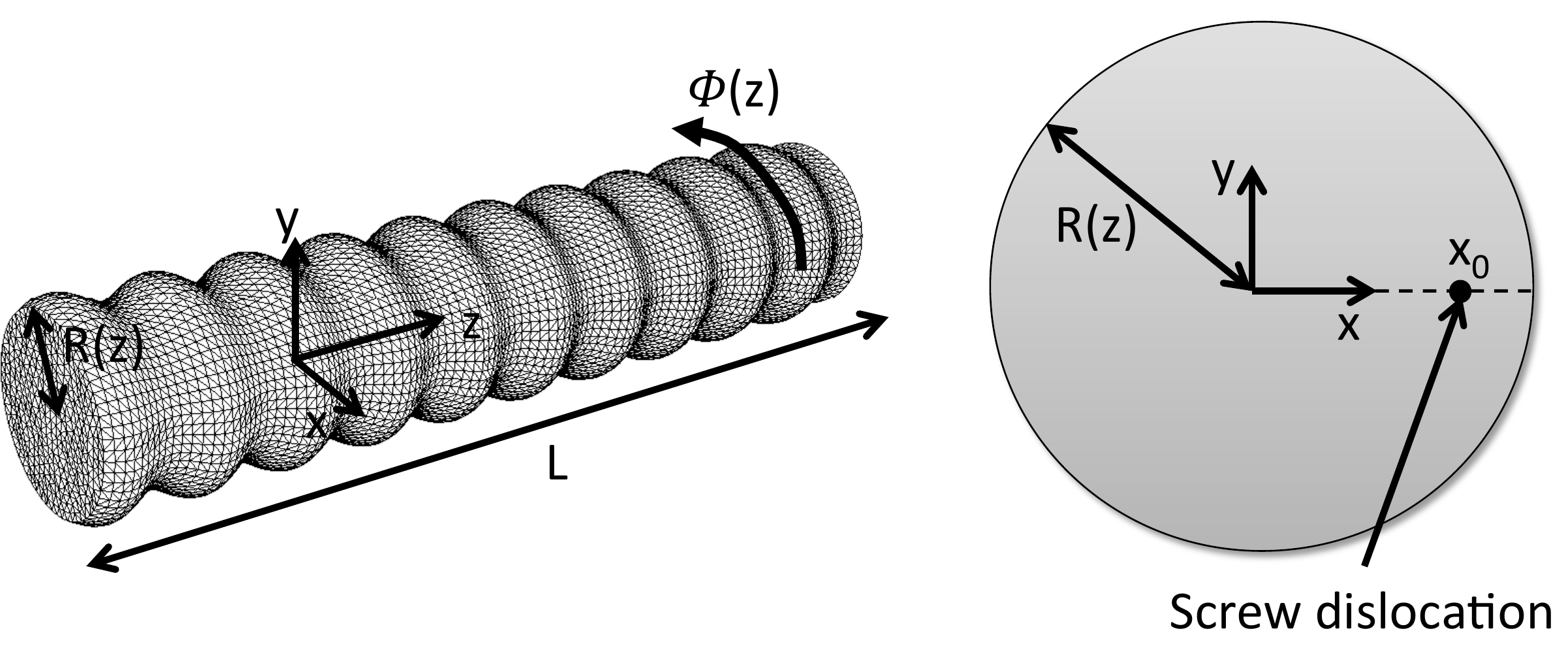}
\caption{Sketch of the geometry of a nanowire. 
A straight screw dislocation is located at $x=x_0$, which is not visible in the left panel.
It shows a FEM discretized nanowire, which contains also a surface corrugation.
Parameters there are $L/R_0=10$, $\delta R/R_0=0.1$ and $kR_0=2\pi$.
We use here 60 nodes in a cross-section and 121 nodes in longitudinal direction.
The right panel is a cross section of the nanowire.
 }
\label{wire::fig1a}
\end{center}
\end{figure}

Let us assume that the nanowire is aligned along the $z$ axis and has constant diameter $R_0$.
A screw dislocation with Burgers vector $b$ is pointing in $z$ direction, at a distance $x_0$ from the centre of the wire, see Fig.~\ref{wire::fig1a}.
%

%In torsion approximation without warp
The displacement field reads
\begin{eqnarray}
u_x &=& -\tau y z, \qquad u_y = \tau x z,  \label{wire::eq1} \\
u_z &=& \frac{b}{2\pi} \left( \arctan \frac{y}{x-x_0} - \arctan \frac{y}{x-R_0^2/x_0} \right). \label{wire::eq3} 
\end{eqnarray}
Here, $\tau$ is the torsion angle per length.
The $z$ component of the displacement field contains as first term the field of a screw dislocation located at $x=x_0, y=0$ in infinite space.
Since boundary conditions demand the absence of tractions on the wire boundary $r=R_0$ (in polar coordinates $r, \theta$), additionally the field of a ``mirror charge'' appears as second term in the expression.
This mirror term is not present in the singular limit $x_0=0$.

The elastic energy $E$ per unit length of the wire follows from integration of the elastic energy density $e=\sigma_{ij}\epsilon_{ij}/2$ as
\begin{equation}
E(x_0, \tau)/L = \frac{1}{4} \mu \tau^2 \pi R_0^4 + \frac{\mu b}{2} (R_0^2 - x_0^2)\tau + \frac{\mu b^2}{4\pi} \ln \frac{R_0^2-x_0^2}{R_0 r_0}.
\end{equation}
Here we have introduced a cutoff radius $r_0$ for the singular core of the screw dislocation.
This energy expression consists of three contributions:
The first term is the energy of the torsion, the last the energy of the screw dislocation, and the second term reflects a coupling between the defect and the torsion.
In particular, without twist, $\tau=0$, the energy is highest for $x_0=0$ and reduced if the dislocation moves closer to the perimeter.

\begin{figure}
\begin{center}
\includegraphics[width=8cm]{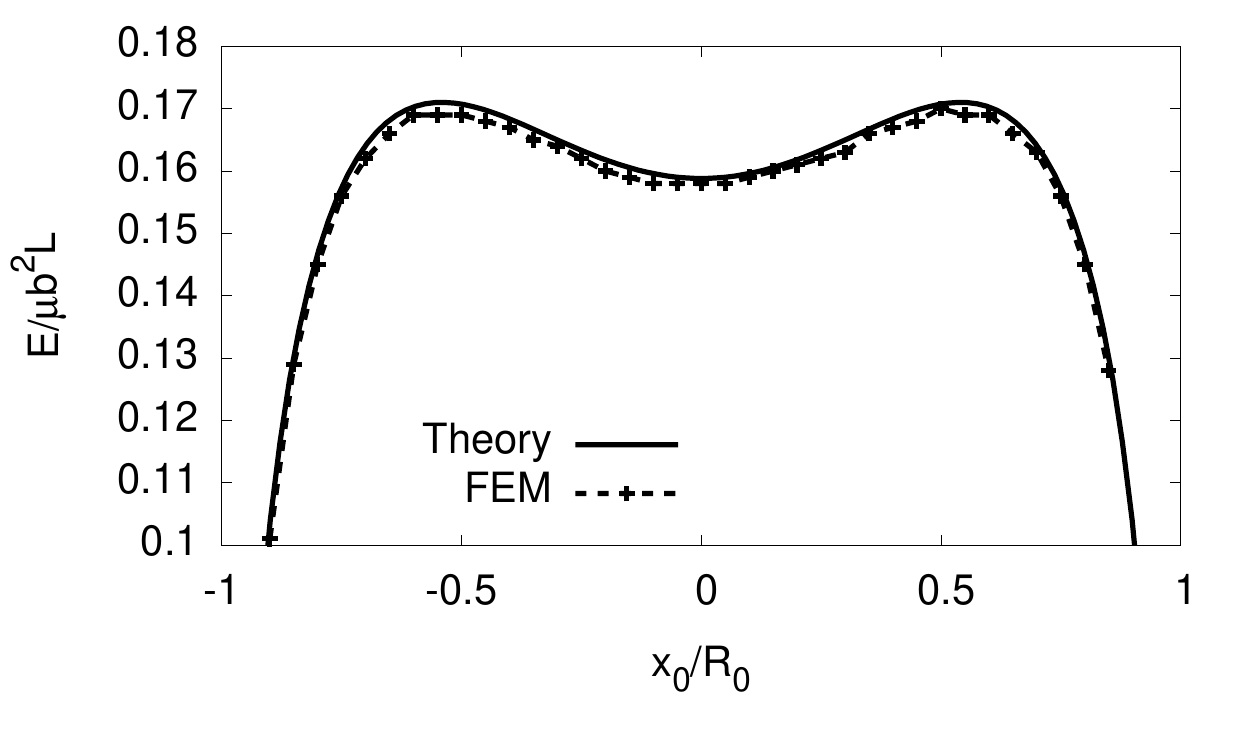}
\caption{Elastic energy per length for a cylindrical nanowire with screw dislocation as function of the position of the dislocation. The centre position $x_0=0$ is metastable.
The graph shows both the analytical theory and the results from finite element simulations, where we use an aspect ratio of $L/R=10$ and a dislocation core radius $r_0/R=0.05$. The wire is discretized by 100 nodes in the circular cross section and 11 nodes in longitudinal direction.}
\label{wire::fig1b}
\end{center}
\end{figure}

So far, we have assumed the torsion angle to be given.
For a long wire with free ends it will adjust such that the energy is minimized, $dE/d\tau=0$.
From this follows the Eshelby twist
\begin{equation} \label{wire::eq5}
\tau(x_0) = - \frac{b}{\pi R_0^4}(R_0^2-x_0^2)
\end{equation}
and for the minimized energy
\begin{equation}
E(x_0)/L = \frac{\mu b^2}{4\pi} \left( \ln\frac{R_0^2-x_0^2}{R_0 r_0} - \frac{(R_0^2-x_0^2)^2}{R_0^4} \right).
\end{equation}
The energy as function of the dislocation position is sketched in Fig.~\ref{wire::fig1b}.
It is reduced in comparison to the case without a spontaneous twist.
Eshelby has derived the same result using a torque balance instead of energy minimisation \cite{Eshelby1953}.
The result is surprising:
The energetic most favourable situation is a case without screw dislocation, as then the elastic energy is zero.
Nevertheless, the centre position $x_0=0$ is metastable, and therefore the dislocation can stay there without being pushed out of the wire by the interaction with the mirror dislocation.

%%%%%%%%%%%%%%%%%%%%%%%%%
\subsection{Torsion of nanowires with non-constant diameter}

In this section we investigate the interplay of torsion of the wire with the initial state of a Rayleigh-Plateau instability.
This instability is driven by a decay of interfacial energy.
Here, however, we will omit the discussion of interfacial contributions and focus on elastic effects, to highlight their role;
the usual interfacial terms can be added to the elastic terms in a straightforward way.
The nanowire is studied here in absence of a screw dislocation, which means that the torsion is imposed by torques acting at the opposite ends of the wire.
We use here an approximative treatment in the framework of the theory of torsion \cite{Landau1986}.
Due to the variation of the wire radius, $R=R(z)$, the absolute torsion angle $\phi$ becomes $z$ dependent.
This angle is related to the torsion per length by $\tau=\phi'(z)$, and intuitively the torsion $\tau$ is larger in the thinner regions of the wire.

The elastic energy of the torsion is given by \cite{Landau1986}
\begin{equation}
E = \frac{1}{2} \int_0^L C(z) [\phi'(z)]^2 dz + E_{ext}
\end{equation}
with the length of the wire $L$ and the external work $E_{ext}$ due to the torques, which are assumed to act only at the faces at $z=0$ and $z=L$.
Here the torsion resistance $C(z)$ is given by
\begin{equation}
C(z) = 4\mu \int (\nabla\chi)^2 df
\end{equation}
which consists of an integral in the plane $z=const$.
The function $\chi$ has to fulfill a Laplace equation, $\nabla^2\chi=0$, inside the wire plane and vanishes on the stress free surfaces, $\chi=0$.
For a cylinder with radius $R(z)$ the solution is
\begin{equation}
C(z) = \frac{\mu\pi R^4(z)}{2}.
\end{equation}
Mechanical equilibrium demands the minimization of energy with respect to the torsion $\phi(z)$.
The variation of energy gives
\begin{equation}
\delta E = - \int_0^L \frac{d}{dz} \left( C(z)\frac{d\phi}{dz} \right) \delta\phi\,dz + \delta E_{ext} = 0,
\end{equation}
where the variation $\delta\phi$ is assumed to vanish at the ends, $\delta\phi(0)=\delta\phi(L)=0$.
Since the external forces act only at the ends, they do not contribute to the bulk variation, and we obtain the generalised torsion condition
\begin{equation}
\frac{d}{dz} \left( C(z)\frac{d\phi}{dz} \right) = 0.
\end{equation}
We therefore get
\begin{equation}
\phi'(z) = C_0 R^{-4}(z),
\end{equation}
in agreement with the expectation that the torsion is larger in the regions with lower diameter.
The integration constant $C_0$ follows from the given total rotation $\phi(0)=0$ and $\phi(L)=\alpha$,
\begin{equation}
C_0 = \alpha \left( \int_0^L R^{-4}(z) dz \right)^{-1}.
\end{equation}
For a sinusoidal perturbation
\begin{equation} \label{wire::eq14}
R(z) = R_0 + \delta R \cos(kz) 
\end{equation}
we obtain for the energy %per period %{\bf (this expression is perhaps incorrect, check it)}
%\begin{equation}
%E = \frac{\mu\pi \alpha^2}{2}  \frac{\sqrt{R_0^2 - (\delta R)^2} (R_0^6 - 3R_0^4 (\delta R)^2 + 3 R_0^2 (\delta R)^4 - (\delta R)^6)}{(3 (\delta R)^2 + 2 R_0^2) R_0 L}.
%\end{equation}
\begin{equation}
E = \frac{\mu\pi \alpha^2}{2}  \frac{\big(R_0^2 - (\delta R)^2\big)^{7/2} }{\big(3 (\delta R)^2 + 2 R_0^2\big) R_0 L}.
\end{equation}
Apparently, the energy vanishes for complete necking, $\delta R=R_0$.
Noticeable, the torsion energy does not depend on the wavelength of the perturbation.
For a small perturbation, $\delta R\ll R_0$, we get in the framework of the torsion theory
\begin{equation} \label{wire::eq16}
E = \frac{1}{4}\mu\pi\alpha^2\left(\frac{R_0^4}{L}-\frac{5 R_0^2}{L}(\delta R)^2\right) + O[(\delta R)^4].
\end{equation}
Apparently, the first term corresponds to the energy of torsion without perturbation.
The second term lowers the energy as a precursor of necking.
Notice that so far interfacial effects are not taken into account in the above expression;
the torsion itself therefore supports the decay into droplets via the curvature driven Rayleigh-Plateau instability.

Finally, up to first order in $\delta R$ the torsion is
\begin{equation} \label{wire::eq17}
\tau(z) = \frac{\alpha}{L} \left( 1 - 4\frac{\delta R}{R_0} \cos kz \right).
\end{equation}

%%%%%%%%%%%%%%%%%%%%
\subsection{Pine-tree nanowires with non-constant diameter}

Here we additionally consider the presence of the screw dislocation in the wire, in combination with a perturbation of the radius, to analyze the interplay between the Eshelby twist and the Rayleigh-Plateau instability.
This problem is treated here in perturbation theory, in a similar spirit as for core-shell nanowires \cite{Wang2008,Schmidt2008}.
The elastic problem is solved via Papkovich-Neuber potentials, and we restrict the analysis to the special case $x_0=0$, i.e. the screw dislocation is sitting in the centre of the nanowire.
More general cases will be discussed below using numerical computations.
Since the calculations are lengthy, we only give essential steps and the final result.

The displacement field is written as
\begin{eqnarray}
\vec{u} &=& 4(1-\nu)\vec{\Psi}-\nabla(\vec{r}\cdot\vec{\Psi} + \Phi), \\
\nabla^2\vec{\Psi} &=& 0, \qquad \nabla^2 \Phi = 0
\end{eqnarray}
with the Poisson ratio $\nu$.

To zeroth order, the fields are given as those of a cylindrical nanowire (constant radius) with a screw dislocation in its centre.
From the above expressions (\ref{wire::eq1}), (\ref{wire::eq3}) follows the stress field in cylindrical coordinates to zeroth order (denoted by the superscript) as
\begin{equation}
\sigma_{rr}^0 = 
\sigma_{\theta\theta}^0 =
\sigma_{zz}^0 = 
\sigma_{\theta r}^0 = 
\sigma_{rz}^0 = 0,\qquad
\sigma_{\theta z}^0 = \frac{1}{2}\frac{\mu b}{\pi r} - \frac{\mu b r}{\pi R_0^2},
\end{equation}
where we used the equilibrium twist according to Eq.~(\ref{wire::eq5}) for $x_0=0$.

Since the main goal is to calculate the elastic energy, which contains the first nontrivial contribution at order $(\delta R)^2$, we determine also the elastic fields up to second order.
For that, the Papkovich-Neuber potentials are written as
\begin{eqnarray}
\Phi(r,z) &=& c_1 I_0(kr)\cos(kz), \\
\Psi_r(r,z) &=& c_2 I_1(kr) \cos(kz), \qquad \Psi_z(r,z) = 0, \\
\Psi_{\theta}(r,z) &=& c_3 I_1(kr) \sin(kz) + 
c_4 I_1(2kr) \sin(2kz) + c_5 I_1(2kr) \cos(2kz),
\end{eqnarray}
with the modified Bessel functions of first kind $I_n$.
The terms involving $2k$ correspond to the second order contributions.
In the end, it turns out that only the first order terms contribute to the elastic energy up to second order.

The coefficients $c_i$ are determined by the stress free boundary conditions at the perimeter of the wire, $\sigma_{ij}n_j=0$.
%Here, the surface normal vector is given by $\vec{n}=n_r \vec{e}_r + n_z\vec{e}_z$ with $n_r = [1+(k\, \delta R)^2 \sin (kz)]^{-1}$ and $n_z = k\,\delta R\, \sin (kz)/[1+(k\, \delta R)^2 \sin (kz)]$.
%From that w
We obtain for the first order expansion coefficients
\begin{equation}
c_1 = 0,\qquad c_2 =0,\qquad c_3= \frac{b\delta R}{8\pi R_0 (\nu - 1)}\frac{1}{I_2(kR_0)}.
\end{equation}
Based on this, also the second order coefficients $c_4$ and $c_5$ can be obtained from the boundary conditions, resulting in lengthy expressions.
Finally, the elastic energy can be calculated from these fields, and it is compared for particular cases in the next section with finite element simulation data.

We can also compare the results from this rigorous perturbation theory with the torsion method from the previous section.
For that, we consider the special case $b=0$, i.e.~the absence of the screw dislocation, and instead a twist which is imposed by an external torque.
First, using the Papkovich-Neuber potentials we obtain for the twist the expression (\ref{wire::eq17}), using the representation $u_\theta=r\phi(z)$.
Next, the stress fields in the framework of the torsion approximation are
\begin{eqnarray}
\sigma_{rr} &=& \sigma_{\theta\theta} = \sigma_{zz} = \sigma_{r\theta} = \sigma_{rz} = 0,\\
\sigma_{z\theta} &=& \mu\tau(z) r = \mu r \frac{\alpha}{L}\left(1-4\frac{\delta R}{R_0}\cos(kz)\right).
\end{eqnarray}
with only one non-vanishing component.
From the rigorous perturbation theory we get
\begin{eqnarray}
\sigma_{rr} &=& \sigma_{\theta\theta} = \sigma_{zz} = \sigma_{rz} = 0, \\
\sigma_{r\theta} &=& -\mu kR_0\tau \frac{I_2(kr)}{I_2(kR_0)}\sin(kz)\delta R, \\ 
\sigma_{z\theta} &=& \mu\tau r -\mu kR_0\tau\frac{I_1(kr)}{I_2(kR_0)}\cos(kz)\delta R \approx \mu r \frac{\alpha}{L}\left(1-4\frac{\delta R}{R_0}\cos(kz)\right),
\end{eqnarray}
where in the last step we used the long wave limit $k R_0\ll 1$, thus $I_1(x)\simeq x/2$ and $I_2(x)\simeq x^2/8$.
Hence in this limit the stress field agrees with the torsion approximation in all components apart from $\sigma_{r\theta}$.
Since this shear component also enters into the expression for the elastic energy, we get from the perturbation theory an expression, which slightly differs from Eq.~(\ref{wire::eq16}),
\begin{equation}
%E_P = \frac{1}{8n}\mu\alpha^2 k \left(R^4 - \frac{5}{9}R^2\delta^2\right).
E = \frac{1}{4}\mu\pi\alpha^2\left(\frac{R_0^4}{L}-\frac{5 R_0^2}{9L}(\delta R)^2\right) + O[(\delta R)^4].
\end{equation}
%{\bf Rewrite this expression in the style of the other.}
We can therefore conclude that the torsion approximation gives qualitatively a correct description, but does not accurately describe the energy of the perturbation.
It is nevertheless useful in particular since it is easy to obtain and works also beyond the limit of small perturbations of the wire diameter, even predicting the necking transition correctly.

%%%%%%%%%%%%%%%%%%%%
\subsection{Finite element modelling of pine-tree nanowires}

A complete analytical treatment of a nanowire with screw dislocation and surface corrugation is not possible if the dislocation line is not in the centre, $x_0\neq 0$.
We therefore use a finite element implementation based on FreeFEM++ to solve the problem numerically \cite{freefem}.

To avoid the appearance of the displacement discontinuity of the dislocation in the weak formulation of the elastic problem we separate this part and treat it analytically.
Hence we write $u_i=u_i^0 + \delta u_i$, where the field $u_i^0$ contains the part described by Eqs.~(\ref{wire::eq1}) and (\ref{wire::eq3}).
From this, the fields $\epsilon_{ij}^0$ are derived from the definition of the strain and $\sigma_{ij}^0$ from Hooke's law.
The weak form of the static elastic equation $\partial\sigma_{ij}/\partial x_j=0$ becomes then after multiplication with a test function $v_i$ and integration over the volume $V$ of the wire
\begin{equation}
\int\limits_V \delta\sigma_{ij} w_{ij} dV + \int\limits_{\partial V_\mathrm{jacket}} \sigma_{in}^0 v_i dS - \int\limits_{\partial V_\mathrm{down}} \delta\sigma_{in}v_i dS + \int\limits_{\partial V_\mathrm{up}} \sigma_{in}^0 v_i dS = 0,
\end{equation}
where we used Gauss' theorem and defined $w_{ij}=(\partial_j v_i+\partial_i v_j)/2$.
We fix here the displacement on the lower surface but let the upper one stress free, such that the Eshelby twist can develop.
The surface $\partial V$ splits into the jacket, $r=R(z)$, upper surface $z=L$ and lower surface $z=0$.
In the bulk, the dislocation core region $(x-x_0)^2+y^2<r_0^2$ is omitted from the volume integral.

Fig.~\ref{wire::fig1b} shows the result of FEM simulations for a nanowire with constant diameter and a straight screw dislocation inside.
With free ends, the wire forms the Eshelby twist as predicted theoretically, and the plot shows the relaxed elastic energy in comparison with the theoretical prediction, demonstrating a good agreement between both of them.

\begin{figure}
\begin{center}
\includegraphics[width=6cm]{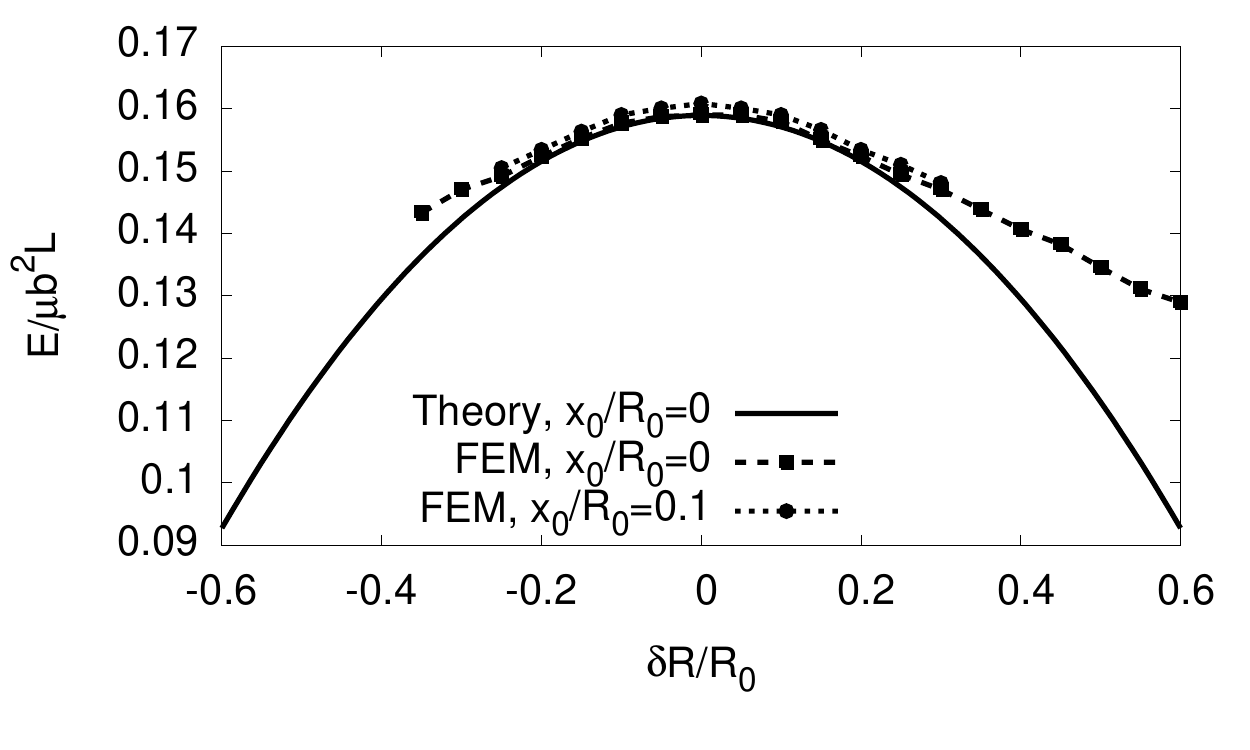}
\includegraphics[width=6cm]{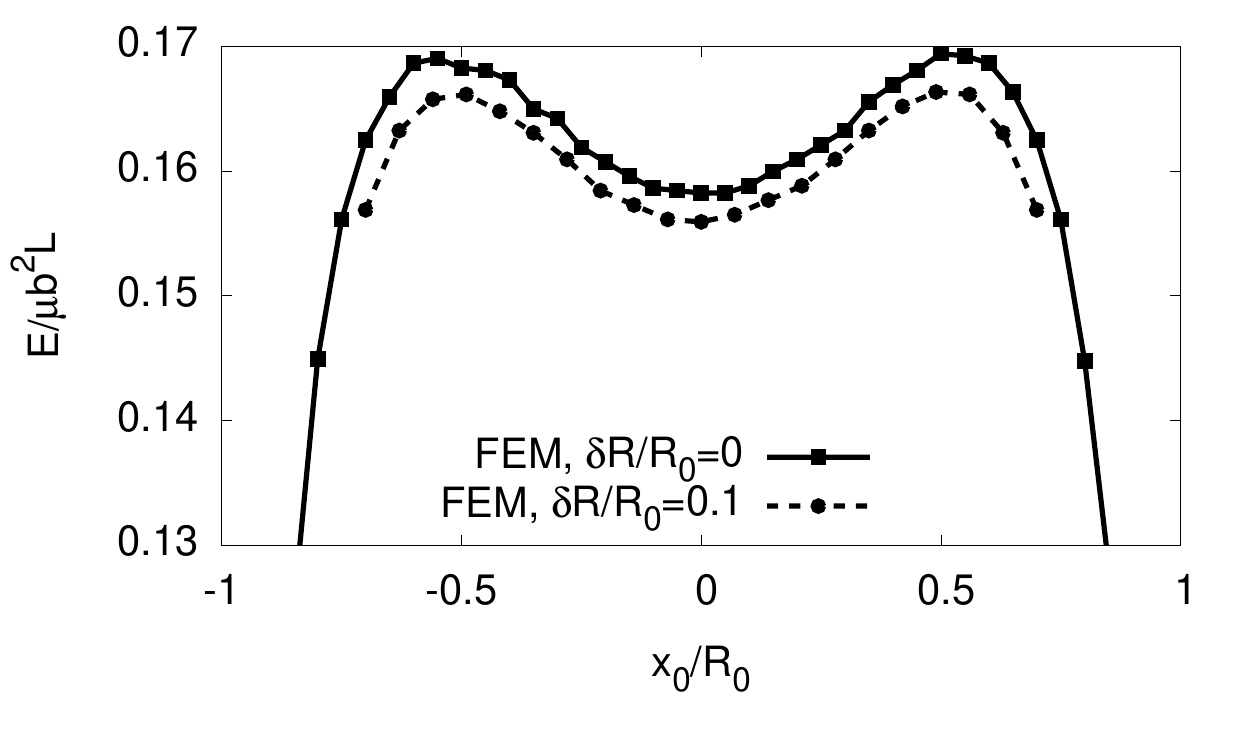}
\caption{Left: Comparison of the elastic energy for a nanowire with perturbed surface, using perturbation theory for $x_0=0$ and two different offset positions using FEM simulations. We use $r_0/R_0=0.05$, $L/R_0=10$, $kR_0=2\pi$ and a discretisation of $60\times 121$ nodes. 
Right: Elastic energy as function of the dislocation core position with ($\delta R/R_0=0.1$) and without perturbation ($\delta R/R_0=0$), for the same parameters as above and a $150\times 101$ mesh.
}
\label{wire::fig3}
\end{center}
\end{figure}

For an additional surface corrugation, $R=R(z)$, the nanowire is discretised appropriately as shown in Fig.~\ref{wire::fig1a}.
%
%\begin{figure}
%\begin{center}
%%\includegraphics[width=12cm]{cone_lambda=1_delta=01.png}
%\includegraphics[width=12cm]{Figures/fig3/fig3.png}
%\caption{Discretized nanowire for FEM simulations with surface perturbation.
%Parameters are $L/R_0=10$, $\delta R/R_0=0.1$ and $kR_0=2\pi$.
%We use here 60 nodes in a cross-section and 121 nodes in longitudinal direction.
%The screw dislocation is not shown.}
%\label{wire::fig2}
%\end{center}
%\end{figure}
%
The elastic energy as function of the perturbation amplitude $\delta R/R_0$ is shown in the left panel of Fig.~\ref{wire::fig3} using the analytical expression obtained from the perturbation theory for $x_0=0$.
It is compared to numerical results for two different positions $x_0$ of the screw dislocation.
At low amplitudes $\delta R/R_0\ll 1$ we find very good agreement between theory and the FEM results.
At larger amplitudes, the decay of the elastic energy is weaker than predicted by perturbation theory. %, which slows down the development of the Rayleigh-Plateau instability.
The dependence of the energy on the position of the screw dislocation is rather weak, in agreement with the knowledge that the centre position $x_0=0$ is a metastable minimum, and therefore changes are only quadratic in $x_0$.
The right panel of Fig.~\ref{wire::fig3} compares the elastic energy of the pine tree nanowire as function of the dislocation position $x_0$ for a situation with and without perturbation.
Here we see that the general structure of the energy landscape does not change by the perturbation, i.e.~the centre position $x_0=0$ remains metastable, but the energy is lowered by the perturbation, which promotes the capillary-driven Rayleigh-Plateau instability.

%%%%%%%%%%%%%%%%%%%%
\subsection{Amplitude equations modelling of pine tree nanowires}

In the descriptions above we focused exclusively on static situations and used the energy to distinguish between stable, metastable and unstable configurations.
In this section we show how the concept of amplitude equations can be used to get insights also into the dynamics of pine tree nanowires.
Here we are interested in particular in the dynamics of the dislocation and the formation of the twist in the nanowire.

Since this -- also in conjunction with the Rayleigh-Plateau instability -- is an interfacial pattern formation problem, the use of phase field methods is suggested.
However, a conventional phase field model may be able to capture elastic effects and defects like cracks \cite{SpatschekHartmann2006,Spatschek11,FleckPilipenk2011,SpatschekBrener2008,Nestler2007}, but a proper representation of dislocations calls for a description using atomic resolution.
Here, during the past years the amplitude equations methods, which can be derived from phase field crystal or classical density functional theory descriptions, have turned out to be useful, and shall therefore be used also for the modelling of pine tree nanowires.
For a thorough introduction into this modelling technique we refer to \cite{spatschek_karma,kar13,Adland:2013ys}.
%The general idea is that time-averaged atomic density is represented as a superposition of density waves,
%\begin{equation}
%\rho(\vec{r}) = \rho_0 \left( 1 + \sum_j A_j(\vec{r}) \exp(i\vec{k}_j\cdot \vec{r}) \right)
%\end{equation}
%with complex amplitudes $A_j$.
%Their complex phase encodes elastic deformations as well as dislocations.
%The sum in the above expression runs on the set of principal reciprocal lattice vectors.
%The amplitudes obey evolution equations of Allan-Cahn type
%\begin{equation}
%\frac{\partial A_j}{\partial t} = -\frac{\delta F}{\delta A_j^*}
%\end{equation}
%$\partial A_j/\partial t = -\delta F/\delta A_j^*$.
%with a free energy functional.
%For bcc materials, this expression has been derived in \cite{Spatschek:2010fk}.

\begin{figure}
\begin{center}
\includegraphics[trim=10cm 0cm 10cm 0cm, clip=true, width=5cm]{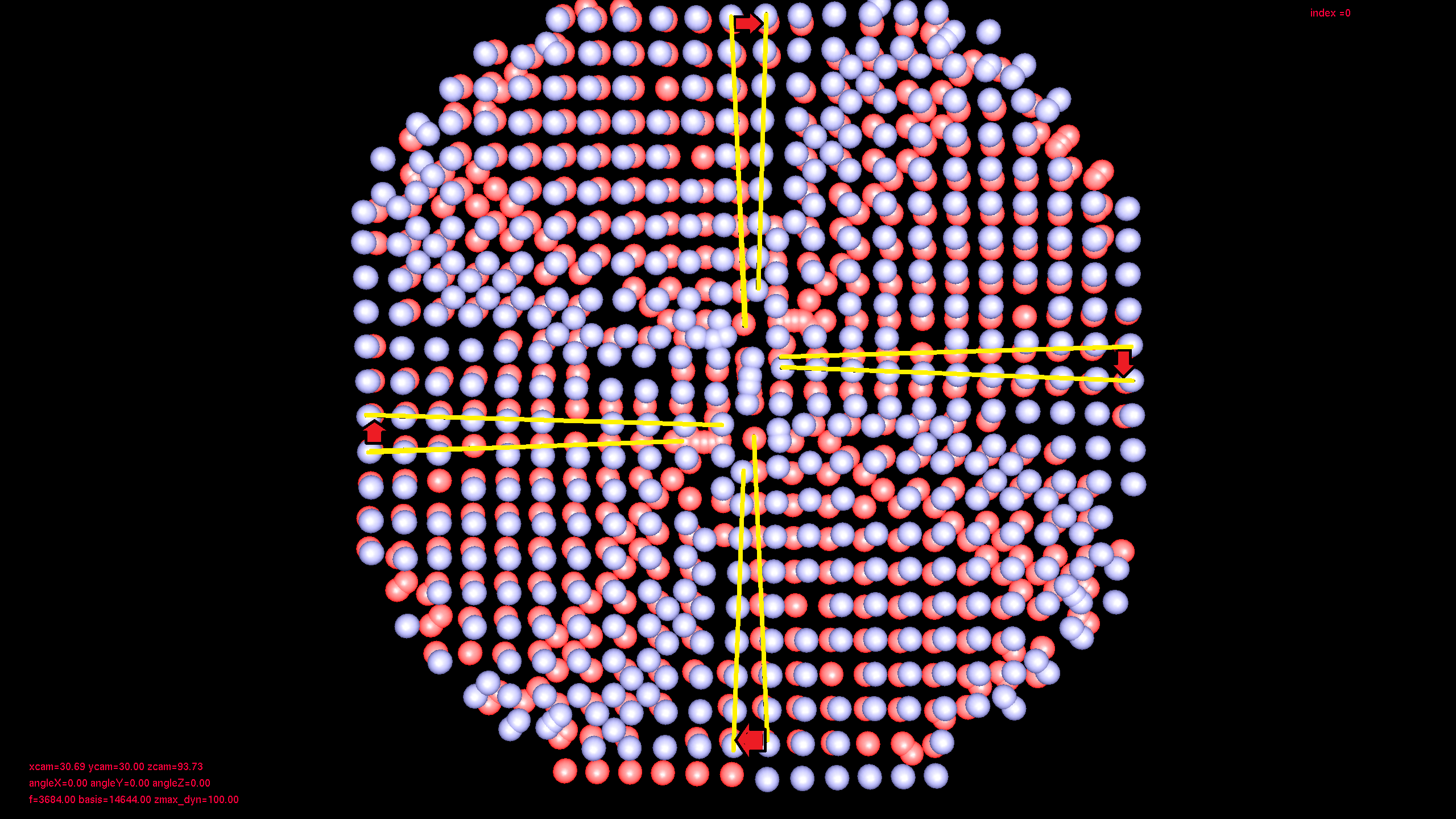}
\caption{Amplitude equations model of a pine tree nanowire. The atomic positions are reconstructed from the amplitudes.
%, visualised here by spherical atoms.
Different colour coding is used for two atomic layers, which are separated by a few lattice units. The upper layer (blue) is rotated clockwise against the lower layer (red) due to the Eshelby twist.}
\label{wire::fig4}
\end{center}
\end{figure}

For the simulations we embed the wire 
%(the domain in space where the amplitudes have magnitude $|A_j|=1$) 
into a liquid or vapour phase.
% with $A_j$=0.
Since at the bulk coexistence point the wire would ``melt'' due to curvature effects, we either have to lower the temperature to stabilise it (we use a Lagrange multiplier to conserve the amount of the solid phase), or we embed the wire into an inhomogeneous temperature field, which is below the melting point inside the wire and above outside.
We use both methods, which both have the advantage that the wire can elastically deform, as the surrounding liquid phase does not support stresses, and therefore the surfaces of the wire are traction free.
As a result, the Eshelby twist builds up in order to minimise the elastic energy, and this is shown in Fig. \ref{wire::fig4}.

\begin{figure}
\begin{center}
\includegraphics[trim=14cm 6cm 14cm 6cm, clip=true, width=4cm]{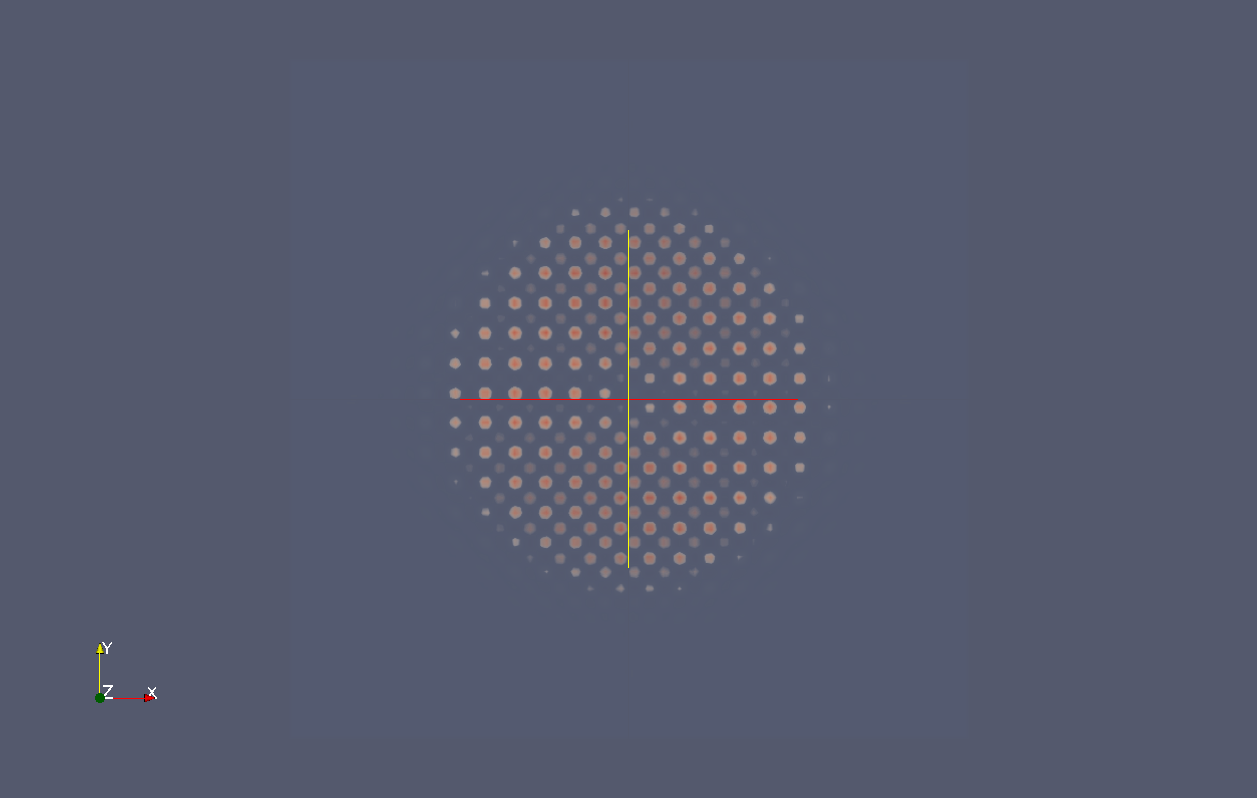}
\includegraphics[trim=14cm 7cm 14cm 5cm, clip=true, width=4cm]{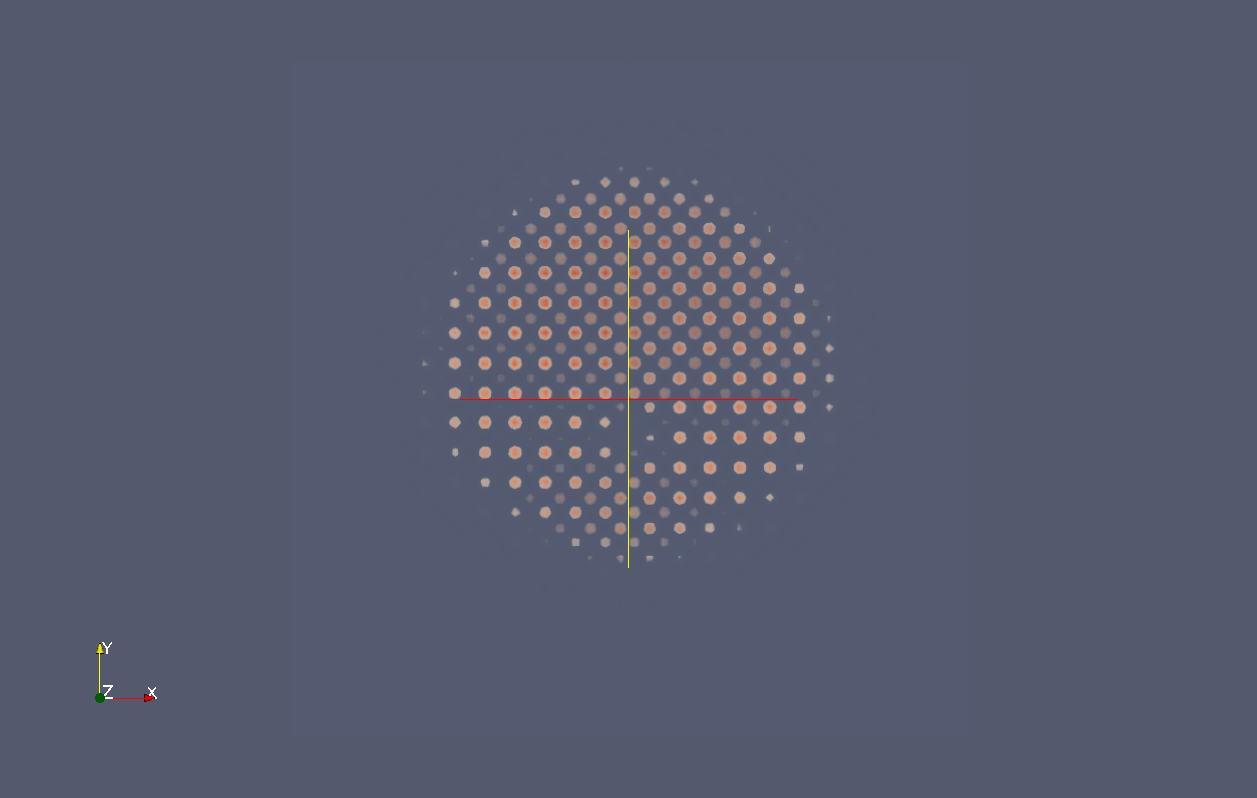}
\includegraphics[trim=14cm 6cm 14cm 6cm, clip=true, width=4cm]{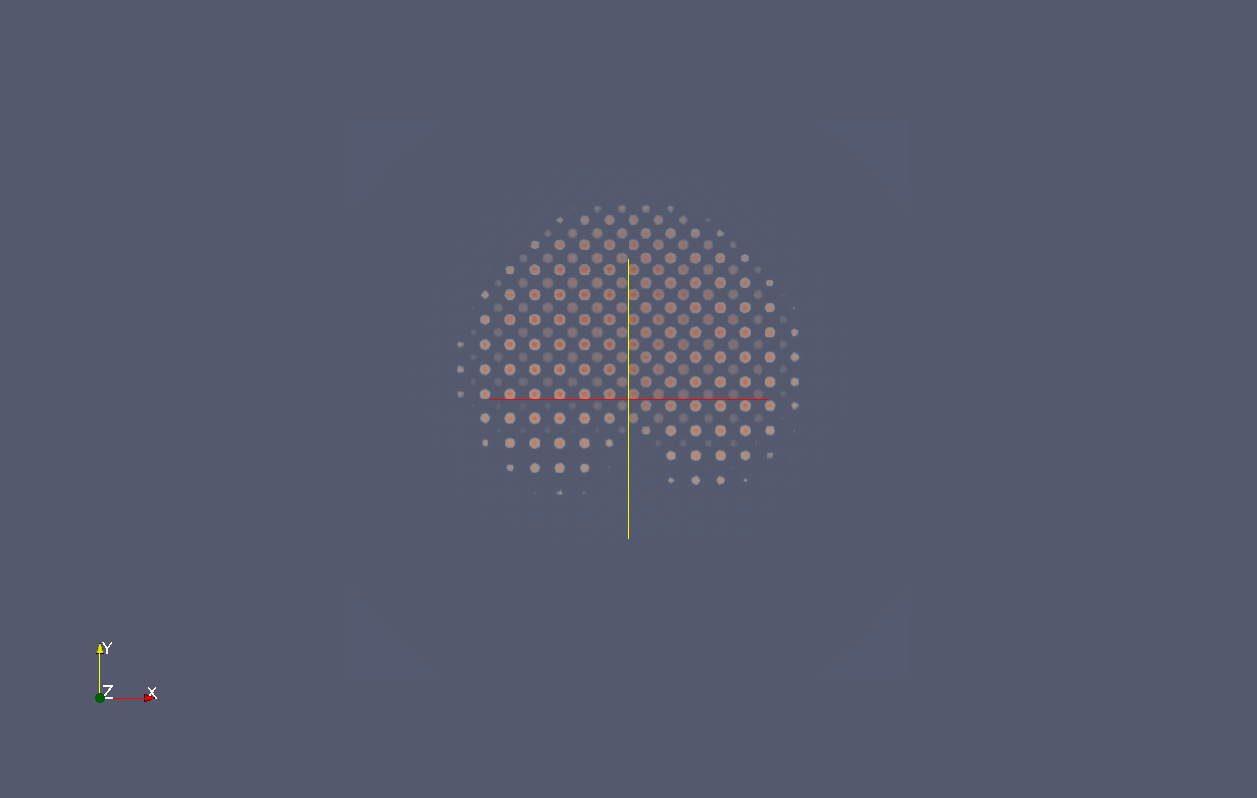}
\caption{Amplitude equations modelling of a pine tree nanowire with fixed ends at $x=0$ and $x=L$, hence the Eshelby twist cannot develop. Then the dislocation does not have a metastable position at $x_0=0$ and therefore drifts out of the wire. The figures show cross sections of snapshots of the temporal evolution. Initially, the screw dislocation is located in the centre and then diffuses out of the wire.}
\label{wire::fig5}
\end{center}
\end{figure}

We can consider also a case where the wire is clamped at its end, and then the Eshelby twist cannot develop.
In this case, the centre position for the dislocation $x_0=0$ is no longer metastable but a maximum of the energy, and therefore a dislocation always drifts out of the wire.
In the amplitude equations model the dislocations are mobile, and this is illustrated in Fig.~\ref{wire::fig5}, which shows how the dislocation is ejected from the nanowire to minimize the elastic energy.

\section{Conclusion}

We have studied theoretically elastic and plastic effects on heterogeneous nucleation and nanowire formation. In the first part of the article, we have investigated the nucleation of a crystalline phase in a confined geometry between two parallel substrates having a lattice misfit with the crystal. We use scaling laws for the elastic energy of the system that are allowing to describe analytically the different nucleation regimes as a function of the thermodynamic driving force for nucleation and the degree of confinement.
From that, we have predicted the corresponding phase diagrams. In the case of no confinement, we have complemented this analytical study with amplitude equations simulations that are supporting the qualitative statements derived from scaling laws, and especially the elongation of the critical nucleus perpendicularly to the substrate with increasing lattice misfit. 

In the second part of the article, we have studied the interplay between Eshelby twist and the Rayleigh-Plateau instability in nanowires. We show  analytically that the torsion, inherited from the presence of the screw dislocation in Eshelby's investigation, promotes the undulation of the surface of the nanowire.
In particular, our analytics predict the necking transition of the nanowire, i.e. the decay of the nanowire into separate droplets driven by elastic effects alone, also with an externally applied torque. We show moreover that the finite-element method and amplitude equations of the phase-field crystal model are efficient numerical tools to tackle this problem. 

The authors acknowledge support by the DFG priority program SPP 1296.

%\begin{table}
%\caption{Please write your table caption here.}
%\begin{tabular}{lll}
%\hline\noalign{\smallskip}
%\end{tabular}
%\end{table}
%

\end{document}